\documentclass{IEEEtran4PSCC}

\usepackage{thmbox} 
\newtheorem[M, bodystyle=\normalfont\noindent]{definition}{Definition}

\newtheorem[M, bodystyle=\normalfont\noindent]{problem}{Problem}

\usepackage[ruled,vlined,shortend, linesnumbered]{algorithm2e} 
\usepackage[]{algorithmic} 
\usepackage{amssymb}

\usepackage[]{caption}
\usepackage{subcaption}
\usepackage{dblfloatfix}

\ifCLASSINFOpdf
   \usepackage[pdftex]{graphicx}
    
\else
   \usepackage[dvips]{graphicx}
\fi
\usepackage{booktabs}
\usepackage[ruled,vlined,shortend, linesnumbered]{algorithm2e} 
\usepackage[]{algorithmic} 

\usepackage[cmex10]{amsmath}
\usepackage{bm}

\usepackage{hyperref}

\usepackage{xcolor}

\makeatletter
\let\old@ps@headings\ps@headings
\let\old@ps@IEEEtitlepagestyle\ps@IEEEtitlepagestyle
\def\psccfooter#1{%
    \def\ps@headings{%
        \old@ps@headings%
        \def\@oddfoot{\strut\hfill#1\hfill\strut}%
        \def\@evenfoot{\strut\hfill#1\hfill\strut}%
    }%
    \def\ps@IEEEtitlepagestyle{%
        \old@ps@IEEEtitlepagestyle%
        \def\@oddfoot{\strut\hfill#1\hfill\strut}%
        \def\@evenfoot{\strut\hfill#1\hfill\strut}%
    }%
    \ps@headings%
}
\makeatother

\psccfooter{%
        \parbox{\textwidth}{\hrulefill \\ \small{24th Power Systems Computation Conference} \hfill \begin{minipage}{0.2\textwidth}\centering \vspace*{4pt} \includegraphics[scale=0.06]{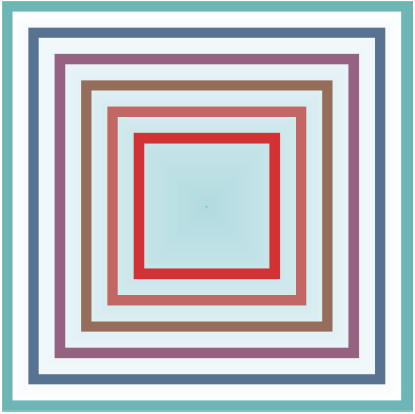}\\\small{PSCC 2026} \end{minipage} \hfill \small{Limassol, Cyprus --- June 8 -- 12, 2026}}%
}

\usepackage{tikz}
\usetikzlibrary{arrows.meta}
\tikzset{>=Stealth}

\usepackage{balance}
\usepackage{easyReview} 

\usepackage{tikz}
\usetikzlibrary{arrows.meta}
\usepackage{circuitikz}

\begin{document}
\title{Multi-Period Sparse Optimization for Proactive Grid Blackout Diagnosis}

\author{Qinghua Ma$^1$,  Reetam Sen Biswas$^2$, Denis Osipov$^3$, Guannan Qu$^4$, Soummya Kar$^4$, Shimiao Li$^1$\\
$^1$EE Dept., University at Buffalo
\{qinghuam, shimiaol\}@buffalo.edu\\
$^2$GE Vernova Advanced Research
\{Reetam.SenBiswas\}@gevernova.com\\
$^3$New York Power Authority
\{denis.osipov\}@nypa.gov\\
$^4$ECE Dept., Carnegie Mellon University
\{gqu, soummyak\}@andrew.cmu.edu\\
}

\maketitle

\begin{abstract}
Existing or planned power grids need to evaluate survivability under extreme events, like a number of peak load overloading conditions, which could possibly cause system collapses (i.e. blackouts). For realistic extreme events that are correlated or share similar patterns, it is reasonable to expect that the dominant vulnerability or failure sources behind them share the same locations but with different severity.  Early warning diagnosis that proactively identifies the key vulnerabilities responsible for a number of system collapses of interest can significantly enhance resilience. This paper proposes a multi-period sparse optimization method, enabling the discovery of {persistent failure sources} across a sequence of collapsed systems with increasing system stress, such as rising demand or worsening contingencies. This work defines persistency and efficiently integrates persistency constraints to capture the ``hidden'' evolving vulnerabilities. Circuit-theory based power flow formulations and circuit-inspired optimization heuristics are used to facilitate the scalability of the method. Experiments on benchmark systems show that the method reliably tracks persistent vulnerability locations under increasing load stress, and solves with scalability to large systems ({on average} taking {around} 200 s per scenario on 2000+ bus systems).
\end{abstract}


\thanksto{\noindent This manuscript has been accepted by the 24th Power Systems Computation Conference (PSCC 2026).}

\begin{figure*}[b]
    \centering
    \begin{subfigure}[b]{1\linewidth}
         \centering
         \includegraphics[width=01\linewidth]{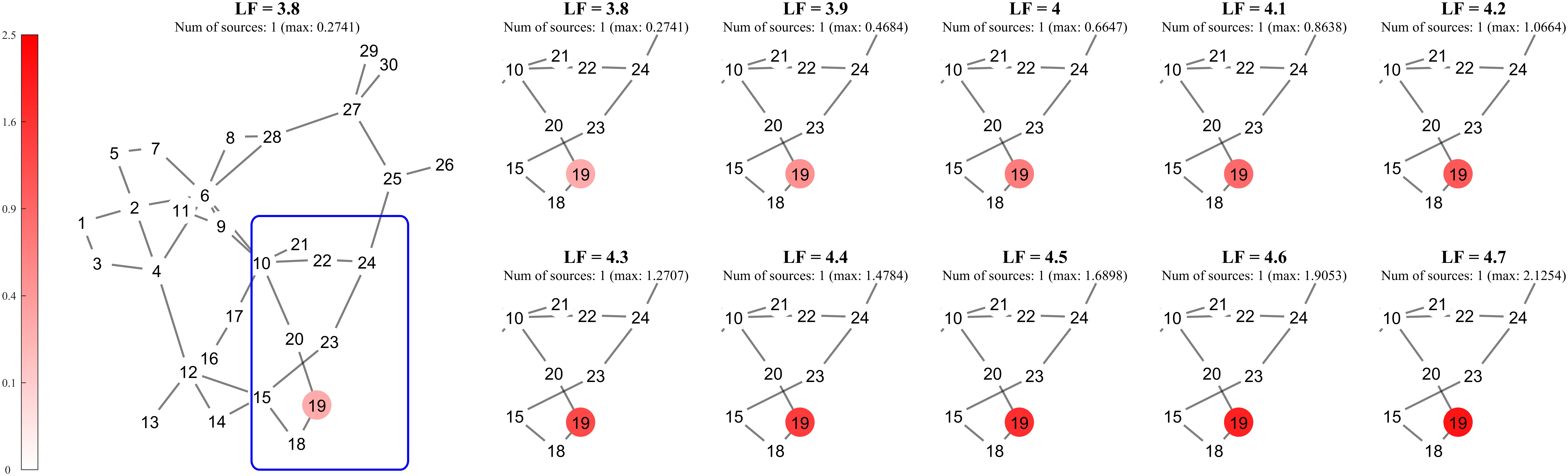}
         \caption{\textsc{Case30}}
         \label{fig: case30 graph}
     \end{subfigure}
     \hfill
       \begin{subfigure}[b]{1\linewidth}
         \centering
         \includegraphics[width=01\linewidth]{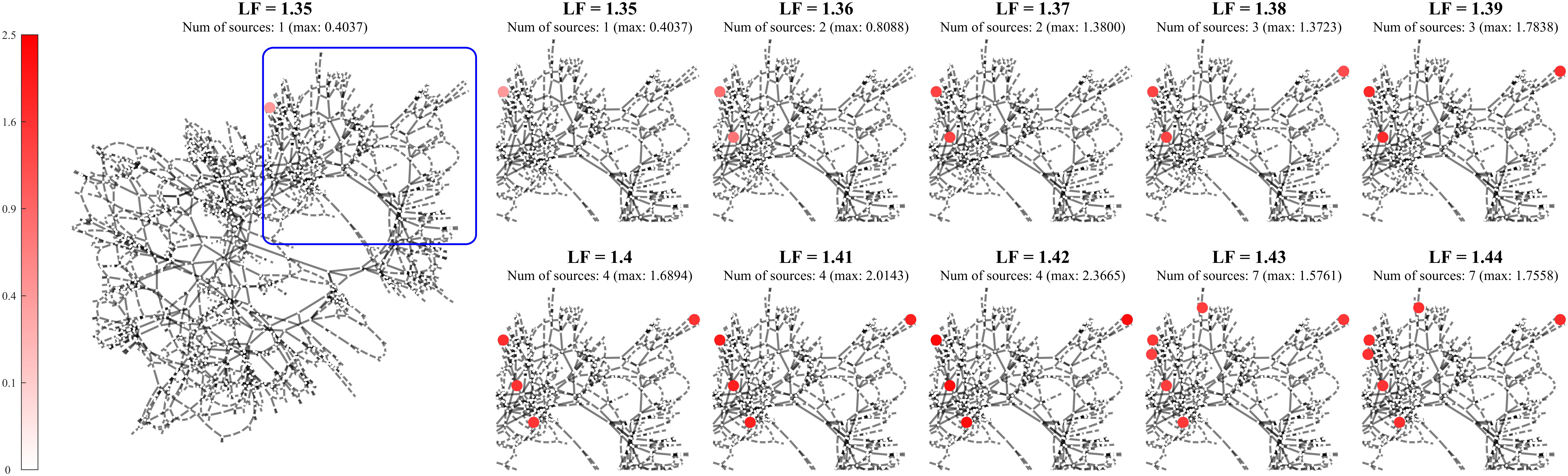}
         \caption{\textsc{Case2383wp}}
         \label{fig: case2383wp graph}
     \end{subfigure} 
    \caption[]{With the growing demand stress, the proposed method identifies a gradually expanding set of system vulnerabilities. Red markers indicate vulnerability locations, with marker size and color representing the quantified severity of vulnerabilities.}
    \label{fig: SparseFeas-MP graph result}
\end{figure*}

\section{Introduction}
\label{sec:Introduction}
The power grid is increasingly challenged by extreme events (e.g., extreme heat or cold) and cyber threats \cite{grid-security-Vyas}, lacking resilience to which can result in system collapses (i.e., blackouts). In June 2025, a severe heatwave pushed New York’s electric grid to its limits, triggering widespread outages across the Tri-State area affecting thousands of Queens residents. The 2021 Texas power crisis \cite{2021TexasPowerCrisis}, caused by extreme cold in winter, left 4.5 million homes without power.  Ukraine Cyber-attacks in 2015-2016 \cite{ukrain2015}\cite{lee2017crashoverride} brought down sections of the Ukrainian power grid causing power outages for roughly 230,000 consumers. New attack models such as MadIoT \cite{madiot}\cite{gridwarm} are theoretically capable of manipulating load demand at a large number of locations to make the system unstable or collapse. 

With the growing threats, power system operators and planners increasingly need to evaluate survivability under a large set of extreme events (e.g., a set of possible peak or abnormal load conditions) which could possibly cause system collapse \cite{powerblackout}. {Early diagnosis that proactively detects system vulnerability and identifies the impact of the failure} can significantly improve reliability and resiliency against high-impact failures. 

Today's grid operators and planners mainly perform such routine evaluations using the industry-standard power flow simulators \cite{powerflow-diverge} that find a feasible solution from steady-state power flow equations. As a collapsed system mathematically represents an infeasible system whose solution does not exist, these steady-state simulators will diverge and thus return limited information. Later works \cite{sugar-pf}\cite{sugar-pf-Amrit}\cite{infeasibility-quantified-pf-trad} introduce slack variables to capture the violation of system feasibility constraints and thus are able to converge with the severity of collapse quantified. More recently, our prior work \cite{SparseFeas} proposed a sparse optimization framework that rapidly identifies a minimal set of critical locations responsible for a given system collapse, offering targeted insights into the dominant sources of failure and locating key system vulnerabilities. Later works have extended this to distribution systems \cite{SparseFeas4DNet}, as well as combined transmission and distribution networks \cite{SparseFeas4TnDNet}.

However, these advancements are designed to evaluate a single scenario case, and fail to consider multiple important scenarios together. In reality, operators and planners are concerned with many extreme situations that a power grid may face, e.g., a series of anticipated load growth over the next 10 years. Since these multiple scenarios are strongly correlated, sorting them in the order of growing stress naturally forms a practical multi-period sequential problem. Note that the term ``multi-period'' in this work does not refer to fixed time steps, but a sequence of events that are correlated by stress levels. In this setting, it is both intuitive and reasonable to expect that a few key vulnerability locations persist across multiple scenarios, with their severity gradually increasing. By identifying and addressing these hidden ``persistent'' vulnerabilities, decision-makers can desirably reduce blackout risks in an effective and targeted way.

This paper extends the state-of-the-art single-scenario works to {multi-period sparse diagnosis}, enabling the discovery of {persistent failure sources} over a sequence of collapsed scenarios triggered by increasing stress, such as rising demand or worsening contingencies. Unlike independently analyzing each single scenario, this paper defines ``persistency'' and formulates a novel \textit{sparse optimization with persistency prior}. In this way, we integrate  temporal persistency constraints on failure source locations to capture the evolving system vulnerabilities hidden behind blackouts. To solve the problem at scale for large systems, we leverage the efficient location-wise sparse optimization technique  \cite{SparseFeas} and propose to enforce persistency using a soft constraint on sparsity coefficients. We also formulate system constraints under circuit-theoretic modeling and adopt circuit-inspired optimization heuristics \cite{sugar-pf-Amrit} to facilitate scalability to large systems. 

Our results {in Figure \ref{fig: SparseFeas-MP graph result}} show that the method reliably tracks {persistent failure locations} under increasing load demand stress, and solves with scalability to large systems (2383 buses {in Figure \ref{fig: SparseFeas-MP graph result}(b)}). These capabilities support focused decision-making to identify and further reduce blackout risk in both operations and planning, enhancing blackout resilience. 

{Findings of this work will also boost planning efficiency through immediate projections. With the underlying vulnerabilities captured with temporal correlations, solutions from selected load-growth scenarios can be interpolated to intermediate cases. For example, in Figure \ref{fig: case30 graph}, node 19 in the 30-bus system is dominant in load factors 3.8 and 3.9, then it can be expected to remain vulnerable for intermediate load factors such as 3.86 even without explicitly solving. In practice, since it is computationally prohibitive to evaluate every critical case, such projections enable efficient assessment across a wide stress range.}

\section{Related Work} \label{sec:relatedwork}

\subsection{Regular power flow analysis for collapsed systems}

As mentioned earlier, the traditional steady-state power flow analysis converges to a feasible solution $\bm{v}^*$ by solving the nonlinear AC power flow equations $\bm{g}(\bm{v})=\bm{0}$ that characterize network balance  under either polar coordinate (where the state variable $v_i$ at each bus $i$ is defined by the voltage magnitude $|V_i|$ and phase angle $\theta_i$) or rectangular coordinate (where $v_i$ is defined by the real and imaginary voltage $v_i^\text{R}$ and $v_i^\text{I}$); however, if no feasible solution exists for the network, the solver diverges, resulting in limited information. To avoid such divergence, approaches \cite{sugar-pf}\cite{infeasibility-quantified-pf-trad} to capture and quantify infeasibility within power flow analysis were developed to quantify the potential infeasibility within the grid. The approach in \cite{sugar-pf}
introduces slack variable $n_i=n_i^{\text{R}}+\mathrm{j}n_i^{\text{I}}$ to represent ``infeasibility'' at each bus $i$. These values represent compensation terms that capture how much additional resource is needed at each bus to maintain the balance conditions of the network. This infeasibility-quantified power flow study is formulated as a non-convex
optimization problem:
\begin{problem}[Dense optimization]
    \begin{equation}
    \min_{\bm{v},\bm{n}} \frac{1}{2} \|\bm{n}\|_2^2, \text{ s.t. } g_i(\bm{v}) + n_i = 0,\,\forall\,i\label{sugar-pf}
\end{equation}
\end{problem}

This formulation can clearly identify a collapsed (i.e., infeasible) system from a feasible one. Besides, convergence with nonzero $\bm{n}$ vector denotes an infeasible system with severity of power flow deficiency quantified by the magnitude of $\bm{n}$.

\subsection{Sparse diagnosis for collapsed systems}

Although advanced power flow studies \cite{sugar-pf}\cite{infeasibility-quantified-pf-trad} above can converge for a collapsed grid, they do not
indicate specific cause of collapse, nor do they detect
locations that are disrupting power 
   system security and
robustness. In most situations, it would be desirable to know
the smallest possible set of dominant nodes that are causing
system collapse with some quantifiable metric.
Accurate and efficient localization of this dominant set of
nodes identifies the deficiency of power (real and reactive) and
highlights some critical locations for special attention in the
planning process. For instance, consider reactive power
planning (RPP) problems \cite{sparsePFref5-planningFACTS4congestion}\cite{sparsePFref7-planningSVC}\cite{sparsePFref8-planningFACTS}\cite{sparsePFref6-planningSVCnTCSC} that aim to find the optimal
allocation of reactive power support through capacitor banks or FACTS devices such as static VAR compensators (SVC). Such problems correspond to finding the sparsest reactive power compensation vector that satisfies system power balance and operation limits in an optimization-based power flow study. However, solving such optimization problems at scale is challenging. Location selection is typically treated by introducing binary variables, leading to the formulation of mixed-integer programming (MIP) that suffers from combinatorial complexity. Numerous nonlinear equality and inequality constraints further enlarge the problem size and increase non-convexity.

To efficiently locate system vulnerabilities and facilitate efficient proactive defense against system collapses, \cite{SparseFeas} introduced an efficient sparse optimization method. The objective of this method is to provide the
minimal set of (sparse) locations, along with quantified slack sources that can be added to these locations, to restore system feasibility.

Given a single scenario $\bm{g}(\bm{v})$, if the system is collapsed, find a sparse vector $\bm{n}$ that restores feasibility, and return the set of locations $\mathcal{S} = \{i\,|\,|n_i|>\varepsilon\}$ ($\varepsilon=10^{-6}$ adopted in this work) recognized as dominant sources of system vulnerability.
\begin{problem} [Sparse Optimization] \label{prob: single scenario sparse diagnosis} 
\begin{align}
    &\min_{\bm{v},\bm{n}} \frac{1}{2}||\bm{n}||_2^2 + \sum_i c_i|n_i|
    \text{, s.t. }
    g_i(\bm{v}) + n_i = 0,\,\forall\,i
    \label{sparse feas}
\end{align}
\end{problem}

Unlike the formulation \eqref{sugar-pf} that only minimizes the L2-norm and leads to slack sources appearing everywhere in the solution, the proposed approach \eqref{sparse feas} localizes non-zero $n_i$ sources at sparse locations using a novel variant of L1-regularization technique. To efficiently enforce sparsity, the method assigns location-wise sparsity coefficient $c_i$ to each location $i$ and adaptively toggles them to reach highly sparse solutions without requiring large coefficient values, thus removing the risk of ill-conditioning and facilitating efficient convergence. \cite{SparseFeas} demonstrated the efficacy of the sparse diagnosis approach in large systems up to the size of Eastern Interconnection network ($>$80,000 buses) and recognized 1 dominant failure source (vulnerability) when load demand increased by $7\%$.

\subsection{Circuit-theoretic modeling of power grid}\label{bkg: ckt formulation}

A circuit-theoretic formulation for power flow and grid
optimization is developed by leveraging heuristics from circuit simulation \cite{sparsePFref15-spice}, enabling the efficient solution of large-scale power system problems. Instead of defining state variables under polar coordinate and writing power balance equations, this framework models each component within the power grid as an equivalent circuit characterized by its current-voltage (I-V) relationship under rectangular coordinate.  The I-V relationships can represent both transmission and distribution grids without loss of generality and existing works have seen its application to power flow analysis \cite{sugar-pf}\cite{sugar-pf-Amrit}, state estimation \cite{SUGAR-SE-Li}\cite{convexSE-LAV-Li}\cite{ckt-SE-TnD}\cite{ckt-GSE}, and optimal power flow analysis \cite{sparsePFref12-SUGAR-opf} etc. Under circuit-theoretic modeling, power system's network constraints are characterized by Kirchhoff's Current Law (KCL) equations using modified nodal analysis. For computational analyticity, the complex functions are split into real and imaginary parts. Figure \ref{fig: circuit model} shows a toy example of a transmission grid bus $i$. 
\vspace{-3mm}
\begin{figure}[h]
	\centering
	\begin{circuitikz}
		
		\ctikzset{
			resistors/scale = 0.6,
			capacitors/scale = 0.6,
			grounds/scale = 0.6
		}

        \begin{scope}[xshift=-0.5cm]
            \draw (0.5,-.5) to [short, -] (0,-0.5);
            \draw (0,-.5) to [short, -] (0,-1);
            \draw[->,thick,yshift=-0.15cm] (0,-.5) -- (0,-1) node[midway,right] {$i_{\text{Load},i}$};
        	
        	\node[draw, minimum width=4mm, minimum height=4mm, align=center] (load) at (0,-1.5) {$L$}; 
            
            \draw (0,-1) -- (load.north);
        	\draw (load.south) -- (0,-2) node[sground]{};
        \end{scope}

        \begin{scope}[xshift=-2cm]
            \draw (2,-.5) -- (0,-.5);
            \draw (1,-.5) to [short, -] (0,-0.5);
    		\draw (0,-.5) to [short, -] (0,-1);
            \draw[->,thick,yshift=-0.15cm] (0,-1) -- (0,-.5) node[midway,right] {$i_{\text{Gen},i}$};
        	
        	\node[draw, minimum width=4mm, minimum height=4mm, align=center] (gen) at (0,-1.5) {$G$}; 
            
            \draw (0,-1) -- (gen.north);
        	\draw (gen.south) -- (0,-2) node[sground]{};
        \end{scope}
        
		\draw[xshift=1cm] (-1,-.5) to [short, o-] (0,-0.5) to [R,-o] (5,-.5);
        \draw[->,thick,xshift=0.2cm] (0.5,-0.5) -- (1.5,-.5)  node[midway,below] {$i_{\text{line}(i,j)}$};
		\draw (0,-.1) node {$v_i^\text{R}+\mathrm{j}v_i^\text{I}$};
		\draw[xshift=1cm] (5,-.1) node {$v_j^\text{R}+\mathrm{j}v_j^\text{I}$};
		\draw[xshift=1cm] (2.5,0) node {$G_s+\mathrm{j}B_s$};
		
		\draw[xshift=2.3cm] (0,-0.5) to [short, -] (0,-0.8) to [C,-] (0,-1.2) node[sground]{};
		\draw[anchor=west,xshift=1cm] (1.5,-1) node {$\mathrm{j}\dfrac{1}{2}B_{sh}$};
		
		\draw[xshift=4.7cm] (0,-0.5) to [short, -] (0,-0.8) to [C,-] (0,-1.2) node[sground]{};
		\draw[anchor=west,xshift=5.4cm] (-0.5,-1) node {$\mathrm{j}\dfrac{1}{2}B_{sh}$};
	\end{circuitikz}
	\caption{Toy example: transmission system bus $i$ is connected with a PV generator $G$, a PQ load $L$, and a transmission line $(i,j)$  characterized by $\pi$ model. Bus $i$ is a PV bus.}
	\label{fig: circuit model}
\end{figure}

Nodal equations at bus $i$ can be written as below. Eq (\ref{eq: bus i real})-(\ref{eq: bus i imag}) enforce the real ($\cdot^\text{R}$) and imaginary ($\cdot^\text{I}$) current balance.  Eq (\ref{eq: bus i Vset}) imposes the PV bus voltage setting point.
\begin{align}
    &\underbrace{G_sv_i^\text{R}-(B_s+\frac{1}{2}B_{sh})v_i^\text{I}-G_sv_j^\text{R}+B_sv_j^\text{I}}_{i^\text{R}_{\text{line}(i,j)}(v_i^\text{R}, v_i^\text{I}, v_j^\text{R}, v_j^\text{I})}\notag\\
    &~~+ \underbrace{\dfrac{P_{\text{L},i} v_i^\text{R}+Q_{\text{L},i} v_i^\text{I}}{(v_i^{\text{R}})^2 + (v_i^{\text{I}})^2}}_{i_{\text{Load},i}^\text{R}(v_i^\text{R}, v_i^\text{I})}
    - \underbrace{\dfrac{P_{\text{G},i} v_i^\text{R}+Q_{\text{G},i} v_i^\text{I}}{(v_i^{\text{R}})^2 + (v_i^{\text{I}})^2}}_{i_{\text{Gen},i}^\text{R}(v_i^\text{R}, v_i^\text{I}, Q_{\text{G},i})} = 0\label{eq: bus i real}
\end{align}
\begin{align}
    &(B_s+\frac{1}{2}B_{sh})v_i^\text{R}+G_sv_i^\text{I} - B_sv_j^\text{R}-G_sv_j^\text{I}\notag\\
    &~~
    + \dfrac{P_{\text{L},i} v_i^\text{I}-Q_{\text{L},i} v_i^\text{R}}{(v_i^{\text{R}})^2 + (v_i^{\text{I}})^2}
    - \dfrac{P_{\text{G},i} v_i^\text{I}-Q_{\text{G},i} v_i^\text{R}}{(v_i^{\text{R}})^2 + (v_i^{\text{I}})^2} = 0\label{eq: bus i imag}
\end{align}
\begin{equation}
    (v_i^{\text{R}})^2 + (v_i^{\text{I}})^2 = |V_i^{\text{set}}|^2\label{eq: bus i Vset}
\end{equation}

This equivalent circuit formulation enables power systems of any size to be efficiently simulated as an equivalent circuit, numerous convergence techniques that were developed for circuit simulation (e.g. SPICE \cite{sparsePFref15-spice}) can be applied \cite{sugar-pf-Amrit}. This paper focuses on analyzing blackout scenarios where network constraints cannot be satisfied at all nodes, in which case we introduce an additional slack current source to each bus.
\section{Multi-period Sparse Diagnosis} \label{sec:method}

\subsection{Persistency metrics}
\textbf{Task Goal:} Given a set of scenarios $s_1,s_2,\cdots,s_T$ sorted in the order of growing stress to form a multi-period setting, our goal is to find the sparsest and optimal $\bm{n}^{(1)},\bm{n}^{(2)},\cdots,\bm{n}^{(T)}$ vectors that can restore feasibility; and meanwhile recognize dominant vulnerability (or failure source) location sets $\mathcal{S}^{(1)}, \mathcal{S}^{(2)},\cdots,\mathcal{S}^{(T)}$ with desired \textit{persistency}. Here, the \textbf{vulnerability location set} $\mathcal{S}^{(t)}=\{i\,|\,|n_i^{(t)}|>\varepsilon\}$ is defined as the minimal set of locations whose compensations can restore feasibility for scenario $s_t$.

The ideal persistency to be encouraged in our optimal solution is defined below:
\begin{definition}[Ideal persistency of vulnerability locations]Under a multi-period setting $t=1,2,\cdots,T$ with growing stress, the vulnerability location trajectory with ideal persistency satisfies:
\begin{equation}
    \mathcal{S}^{(1)}\subseteq \mathcal{S}^{(2)}\subseteq\cdots\subseteq\mathcal{S}^{(T)}
\end{equation}
Equivalently, a vulnerability location $i$ with ideal persistency satisfies:
\begin{equation}
   \text{if } i\in\mathcal{S}^{(t)}\,\Rightarrow\,i\in\mathcal{S}^{(t+1)},\, \forall\,t\leq T-1\notag
\end{equation}
or, in terms of the sparse indicator:
\begin{equation}
    \text{if } |n_i^{(t)}|>\varepsilon\,\Rightarrow\,|n_i^{(t+1)}|>\varepsilon,\,\forall\,t\leq T-1 \notag
\end{equation}
This intuitively means that once a location shows up as vulnerable, it stays vulnerable in all subsequent scenarios, revealing a critical weakness of the system as the stress keeps increasing.
\label{dfn: ideal persistency}
\end{definition}

However, when analyzing multiple scenarios in practice, the solution may not always satisfy ideal persistency, motivating us to define metrics to quantify the level of persistency achieved. We first define the location persistency metric:
\begin{definition}[Location Persistency Metric] \label{def: location persistency}For a location $i$ under the multi-period setting $t=1,2,\dots,T$, let
$k$
denote the first scenario index at which $i$ becomes a vulnerability location, i.e., $k = \min \{\, t\,|\,i \in \mathcal{S}^{(t)},\; t \leq T-1 \,\}$. 
Firstly, if no such $k$ exists, we set $\mathrm{Persistency}(i)=0$, representing that location $i$ is never vulnerable. 
Secondly, if location $i$ fails to remain vulnerable for at least 2 consecutive steps, we also set $\mathrm{Persistency}(i)=0$, representing that vulnerability at $i$ occasionally occurs but never persists. 
In all remaining cases, we define the \emph{persistency of $i$} as the proportion of subsequent time steps in which
$i$ continues to be vulnerable after it first showed up:
\begin{equation}
    \mathrm{Persistency}(i)= \frac{\sum_{t=k}^T \mathbf{1}_{\{\, i \in \mathcal{S}^{(t)} \,\}}}{T-k+1}{\times100\%}
\end{equation}
where $\mathbf{1}_{\{\cdot\}}$ is the indicator function, equal to $1$ if the condition $\{\cdot\}$ is satisfied and $0$ otherwise. 
This metric satisfies:
\begin{enumerate}
    \item $0 \leq \mathrm{Persistency}(i) \leq 100\%$, $\forall\,i$.
    \item $\mathrm{Persistency}(i) = 100\%$ if and only if $i$ has ideal persistency, i.e., it remains vulnerable at all scenarios $k+1,\cdots,T$ after first showing up at step $k$.
\end{enumerate} 
\end{definition}

Besides location persistency, we further quantify the network-level persistency of all vulnerability locations by introducing the following notion of set persistency:
\begin{definition}[Set Persistency Metric] \label{def: set persistency}From scenario $s_1$ to any scenario $s_t$, we are able to quantify how well the ideal persistency $\mathcal{S}^{(1)}\subseteq \mathcal{S}^{(2)}\subseteq\cdots\subseteq \mathcal{S}^{(t)}$ has been enforced so far, by:
\begin{equation}
    \mathrm{Set Persistency(t)} = \frac{|\mathcal{S}^{(t)}|}{|\cup_{j=1}^t \mathcal{S}^{(j)}|} \times100\%
\end{equation}
where $|\mathcal{S}|$ denotes set cardinality, i.e., the number of elements in set $\mathcal{S}$; and $\cup_{j=1}^t \mathcal{S}^{(j)}=\mathcal{S}^{(1)}\cup \mathcal{S}^{(2)}\cup\cdots\cup\mathcal{S}^{(t)}$ denotes the union of all vulnerability sets so far, which corresponds to a set of all vulnerability locations identified across $s_1,s_2,\cdots,s_t$. This metric satisfies:
\begin{enumerate}
    \item $0 \leq \mathrm{SetPersistency}(t) \leq 100\%$, $\forall\,t$.
    \item When ideal persistency holds, we will have $\mathcal{S}^{(t)}=\cup_{j=1}^t \mathcal{S}^{(j)}$, i.e., $\mathrm{Set Persistency(t)}=100\%$.
\end{enumerate}  
\end{definition}

\subsection{Method overview}

With the task goal and preliminaries defined above, this paper will model the inter-dependence across the multiple scenarios analyzed.

\begin{figure}[h]
\centering
\begin{tikzpicture}[>=Stealth]
\draw[shape=circle]
    (0,0)node(theta1)[draw]{$\bm{\theta}^{(1)}$}
    (1.5,0)node(theta2)[draw]{$\bm{\theta}^{(2)}$}
    (3,0)node(theta3)[draw]{$\bm{\theta}^{(3)}$}
    (4.5,0)node(theta4){$\cdots$}
    (6,0)node(theta5)[draw]{$\bm{\theta}^{(T)}$}
    (0,-1.5)node(x1)[draw]{$\bm{x}^{(1)}$}
    (1.5,-1.5)node(x2)[draw]{$\bm{x}^{(2)}$}
    (3,-1.5)node(x3)[draw]{$\bm{x}^{(3)}$}
    (4.5,-1.5)node(x4){$\cdots$}
    (6,-1.5)node(x5)[draw]{$\bm{x}^{(T)}$};
\draw[->] (theta1.east) -- (theta2.west);
\draw[->] (theta2.east) -- (theta3.west);
\draw[->] (theta3.east) -- (theta4.west);
\draw[->] (theta4.east) -- (theta5.west);
\draw[->] (theta1.south) -- (x1.north);
\draw[->] (theta2.south) -- (x2.north);
\draw[->] (theta3.south) -- (x3.north);
\draw[->] (theta5.south) -- (x5.north);
\end{tikzpicture}
\caption{Bayesian network illustration of multi-period model.}
\label{fig: graphical model}
\end{figure}

For any scenario $s_t$, let the hidden variable $\bm{\theta}^{(t)}=[\theta_1^{(t)}, \cdots,\theta_i^{(t)},\cdots,\theta_N^{(t)} ]$  be a vector of status variables  representing whether each bus location $i\in\{1,2,\cdots,N\}$ is a vulnerability source, i.e., $\theta_i^{(t)}=1$ if $i\in \mathcal{S}^{(t)}$ or equivalently if $|n_i^{(t)}|>\varepsilon$. 

Instead of treating each scenario independently, we adopt a Bayesian Network in Figure \ref{fig: graphical model} as a conceptual lens to capture how hidden vulnerabilities evolve across stress levels, with the sparse optimization output of each scenario serving as a ``noisy observation'' of these hidden vulnerabilities. Specifically, the system starts with an initial hidden vulnerability $\bm{\theta}^{(1)}$. At any time $t$, the change in system stress drives the vulnerability to change incrementally from $\bm{\theta}^{(t-1)}$ to $\bm{\theta}^{(t)}$. And each vulnerability stage $\bm{\theta}^{(t)}$ leads to its corresponding state variables $\bm{x}^{(t)}=[\bm{n}^{(t)}, \bm{v}^{(t)}]$ that include the targeted compensations $\bm{n}^{(t)}$ needed to restore its feasibility and the voltages $\bm{v}^{(t)}$ of the feasible ``reconstructed'' system. The sparse diagnosis method can be considered a sensor or lens from which we can  obtain $\bm{n}$ as a ``tentative observation'' of hidden variables $\bm{\theta}$ and thus locate the hidden vulnerabilities accordingly. 

Under this view, vulnerabilities $\bm{\theta}^{(1)},\cdots,\bm{\theta}^{(t)}$ and outputs $\bm{x}^{(1)},\cdots,\bm{x}^{(t)}$ across all scenarios can be considered jointly, and the Bayesian factorization highlights that the key components are the transition probabilities $P(\bm{\theta}^{(j)}|\bm{\theta}^{(j-1)})$ and the likelihood terms $P(\bm{x}^{(j)}|\bm{\theta}^{(j)})$:
{\small
\begin{align}
\small
    &\bm{\theta}^{(1)*},\cdots,\bm{\theta}^{(t)*}, \bm{x}^{(1)*},\cdots,\bm{x}^{(t)*}\notag\\
    =& \arg\max P(\bm{\theta}^{(1)},\cdots,\bm{\theta}^{(t)}, \bm{x}^{(1)},\cdots,\bm{x}^{(t)}) \notag\\
    =& \arg\max P(\bm{\theta}^{(1)})P(\bm{x}^{(1)}|\bm{\theta}^{(1)})\cdot 
    \prod_{j=2}^t P(\bm{\theta}^{(j)}|\bm{\theta}^{(j-1)})P(\bm{x}^{(j)}|\bm{\theta}^{(j)})
\end{align}}

Considering a multi-period processing in the forward direction, the task above can be split into two subtasks:
\begin{enumerate}
    \item Conduct sparse optimization with persistency prior: For each scenario $s_t$ ($t\geq2$), given $\bm{\theta}^{(t-1)}$ from previous stage, the problem reduces to a single-scenario optimization augmented by the prior probability $P(\bm{\theta}^{(t)}|\bm{\theta}^{(t-1)})$:
    \begin{equation}
    \small
        \bm{\theta}^{(t)*}, \bm{v}^{(t)*}, \bm{n}^{(t)*} = \arg\max P(\bm{\theta}^{(t)}|\bm{\theta}^{(t-1)})P(\bm{v}^{(t)}, \bm{n}^{(t)}|\bm{\theta}^{(t)})
        \label{prob: Bayesian Network Problem t}
    \end{equation}
    \item Obtain the prior distribution $P(\bm{\theta}^{(1)})$ for $t=1$, so that $\bm{\theta}^{(1)*}, \bm{v}^{(1)*}, \bm{n}^{(1)*}$ can be optimized similarly as above.
\end{enumerate}

In practice, however, the exact distribution functions are only partially tractable or entirely intractable (although they theoretically exist), making exact inference from the full joint distribution impractical. Therefore, the remainder of this section develops sparse optimization as a deterministic relaxation of the Bayesian inference problem, while still embedding the notion of vulnerability evolution into our framework.

\subsection{Sparse optimization with persistency prior}
Without prior information on $\bm{\theta}^{(t)}$ for scenario $s_t$, one may assume that every location has the same probability of being a dominant vulnerability source; in which case solving the single-scenario sparse optimization in Problem \ref{prob: single scenario sparse diagnosis} can be seen as finding the most possible $\bm{\theta}^{(t)*}, \bm{v}^{(t)*}, \bm{n}^{(t)*} = \arg\max P(\bm{\theta}^{(t)})P(\bm{v}^{(t)},\bm{n}^{(t)}|\bm{\theta}^{(t)})$ with prior Bernoulli distribution $P({\theta}_i^{(t)}=1)=p$ for $\forall\,i$, where $p$ is a constant.

Now, given prior knowledge $\bm{\theta}^{(t-1)}$ from the previous stage $s_{t-1}$, we can model and integrate a more informative prior distribution $P(\bm{\theta}^{(t)}|\bm{\theta}^{(t-1)})$ to encourage the ideal persistency on stage $s_t$, which, based on our definition, can be illustrated as a state transition probability $P({\theta}_i^{(t)}=1|{\theta}_i^{(t-1)}=1)=100\%$.  This equivalently represents some additional hard constraints on the binary status variables ${\theta}_i^{(t)}$ so that the problem $\eqref{prob: Bayesian Network Problem t}$ becomes:
\begin{align}
    &\min_{\bm{v}^{(t)},\bm{n}^{(t)}} \frac{1}{2}||\bm{n}^{(t)}||_2^2 + \sum_i c_i^{(t)}|n_i^{(t)}|\\
    \text{ s.t. }
    &g_i^{(t)}(\bm{v}^{(t)}) + n_i^{(t)} = 0, \forall\,i \notag\\
    &\text{additional constraint (a): } \theta_i^{(t)}=\mathbf{1}_{\{|n_i^{(t)}|>\varepsilon\}}, \forall\,i\notag\\
    &\text{additional constraint (b): } \theta_i^{(t)}\geq \theta_i^{(t-1)}, \forall\,i \notag
    \label{sparse feas-mp mixed integar}
\end{align}
where (a) defines the binary status variable, and (b) enforces the ideal persistency for each location.

The above problem is mathematically an MIP, which can be hard to solve due to its combinatorial nature. If solution at $s_{t-1}$ is available, one might attempt to remove the mixed-integer variables by equivalently transforming constraints (a) and (b) into $|n_i^{(t)}|> \varepsilon$, for $\forall\,i\in \mathcal{S}^{(t-1)}$, which is a nonlinear and nonconvex hard constraint since $|n_i^{(t)}|=\sqrt{((n_i^{\text{R}})^{(t)})^2+((n_i^{\text{I}})^{(t)})^2}$.

However, practical optimization heuristics that can efficiently enforce sparsity-inducing hard constraints are not readily available. Moreover, ideal persistency may not always be achievable in realistic systems, and enforcing constraint (b) rigidly can lead to no solution or suboptimal solutions with sparsity degraded.

To more efficiently solve the multi-period sparse optimization, we propose to transform the above problem into an equivalent form where we get rid of mixed integers and encourage ideal persistency as a soft constraint. The new problem is defined below:

\begin{problem} [Sparse Optimization with persistency prior] \label{prob: multi-period sparse diagnosis}
\begin{align}
    &\min_{\bm{v}^{(t)},\bm{n}^{(t)}} \frac{1}{2}||\bm{n}^{(t)}||_2^2 + \sum_i c_i^{(t)}|n_i^{(t)}|\\
    \text{ s.t. }
    &g_i^{(t)}(\bm{v}^{(t)}) + n_i^{(t)} = 0, \forall\,i\notag\\
    &\text{sparsity coefficients: } c_i^{(t)}\leq c_i^{(t-1)}, \forall\,i \notag
    \label{sparse feas}
\end{align}
\end{problem}
where we embed ideal persistency into the tuning of sparsity coefficients in each scenario. 

Leveraging the efficient location-wise sparsity enforcement mechanism proposed in \cite{SparseFeas}, where the sparsity coefficients $c_i^{(t)}$ are iteratively and adaptively switched between $c_\text{H}=10$ and $c_\text{L}=0.1$ to promote sparse solutions $\bm{n}^{(t)}$ for each scenario $s_t$. At each update, locations with high $n_i^{(t)}$ values are assigned with $c_\text{L}$; whereas those with lower $n_i^{(t)}$ are assigned with $c_\text{H}$. Upon convergence, $c_i^{(t)}=c_\text{H}$ promotes zero $|n_i^{(t)}|$ and $c_i^{(t)}=c_\text{L}$ promotes non-zero $|n_i^{(t)}|$.

Therefore, we have the sparsity coefficients $c_i^{(t)}$ equivalently serving as the status indicator upon convergence, i.e., $c_i^{(t)}=c_\text{H}$ corresponds to $\theta_i^{(t)}=0$ and $c_i^{(t)}=c_\text{L}$ corresponds to $\theta_i^{(t)}=1$. As a result, the ideal persistency constraint $\theta_i^{(t)}\geq\theta_i^{(t-1)}$ can be transformed into $c_i^{(t)}\leq c_i^{(t-1)}$, and thus embedded into the coefficient toggling process. Algorithm \ref{alg: SparseFeas-MP} shows the pseudocode of our implementation. After each update on the sparsity coefficients, we solve the sparse optimization problem using the circuit formulation as illustrated in Section \ref{bkg: ckt formulation}, that leverages circuit-inspired optimization heuristics \cite{sugar-pf} to facilitate fast convergence. 

\begin{algorithm}
\caption{Sparse optimization with persistency prior for scenario $s_t$ }  
	\label{alg: SparseFeas-MP}
	\KwIn{Prior sparsity coefficients $\bm{c}^{(t-1)}$;  case data $\bm{g}(\cdot)$ for scenario $s_t$}
	\KwOut{$\bm{c}^{(t)}, \bm{n}^{(t)}, \bm{v}^{(t)}$}
	{\bf Initialize: }$c_i^{(t)}=c_\text{H}$ for $\forall\,i$ location 

    {\bf Initialize $ \bm{n}^{(t)}$} by Problem \ref{sugar-pf} dense optimization.

\Repeat{fail to converge to sparser solution}{
Update $\bm{c}^{(t)}$ from  $\bm{n}^{(t)}$ to set a sparser goal (algorithm in \cite{SparseFeas})\;
            \textbf{Encourage persistency} $\bm{c}^{(t)}\preccurlyeq  \bm{c}^{(t-1)}$:\
        $$c_i^{(t)}\gets 0.5\cdot c_\text{L} , \text{for } \forall\,i \text{ with } c_i^{(t-1)}\leq c_\text{L}$$
        
        Solve sparse optimization with updated $\bm{c}^{(t)}$: $\min_{\bm{v}^{(t)},\bm{n}^{(t)}} \frac{1}{2}||\bm{n}^{(t)}||_2^2 + \sum_i c_i^{(t)}|n_i^{(t)}|$ 
        \\[0.5em]s.t. 
        $g_i^{(t)}(\bm{v}^{(t)}) + n_i^{(t)} = 0,\,\forall i$ 
}

\end{algorithm}

\section{Experiments} \label{sec:experiments}
\label{sec:Results}

\noindent This section conducts experiments to evaluate the efficacy of our proposed multi-period method, by answering the following questions:
\begin{enumerate}
    \item \textbf{Q1. Persistency}: how persistent are the identified vulnerability locations from our proposed method compared to
  the baseline method \cite{SparseFeas}?
    \item \textbf{Q2. Sparsity \& total compensations}: do we sacrifice sparsity and total compensations while enforcing persistency?
    \item \textbf{Q3. Scalability}: how does our algorithm scale with the system size?
\end{enumerate}

We test and compare two methods: 
\begin{itemize}
    \item \textbf{(Proposed) Multi-period method}  in Section \ref{sec:method};
    \item \textbf{(Baseline) Single-scenario method }\cite{SparseFeas} that solves Problem \ref{prob: single scenario sparse diagnosis} separately for each scenario.
\end{itemize}

This paper tests on standard IEEE cases of transmission systems. For each test case, we generate $T=10$ synthetic collapsed scenarios by increasing the load factor incrementally. Load factor is the scaling factor used to increase the original power load. Notably, the proposed work itself is agnostic to demand patterns. Sections \ref{sec: persistency result}-\ref{sec: scalability result} mainly focus on uniformly growing load factors across the entire grid,  whereas Section \ref{sec: extended experiments} applies the method to uneven load-growth cases where some locations/areas grow faster than others.

In Section \ref{sec: persistency result}, we show results of \textsc{case30} and \textsc{case2383wp} \cite{matpower-case30-118-300-2383wp}, evaluating location persistency (Definition \ref{def: location persistency}) and set persistency (Definition \ref{def: set persistency}) of  baseline method and proposed multi-period method. In Section \ref{sec: sparsity and total compensation}, we compare their sparsity and total compensation. In Subsection \ref{sec: scalability result}, test cases of different sizes are used to evaluate run time and scalability. Section \ref{sec: extended experiments} further demonstrates the effectiveness of the proposed method under uneven load-growth scenarios. All experiments were conducted on a Microsoft Windows 11 desktop with 3.6~GHz Intel i7-12700K Core and 32 GB RAM. 

\subsection{Persistency analysis} \label{sec: persistency result}
We show results on IEEE standard 30-bus case \textsc{case30} with load factors $3.8, 3.9, \cdots, 4.7$, and \textsc{case2383wp}, which is a 2383-bus case representing part of the European high voltage network, with load factors $1.35, 1.36, \cdots, 1.44$. Figure~\ref{fig: SparseFeas-MP graph result} provides a graphical illustration of the results. 

\begin{figure}[t]
    \centering
    \begin{subfigure}[b]{\linewidth}
    \centering         
         \includegraphics[width=\linewidth]{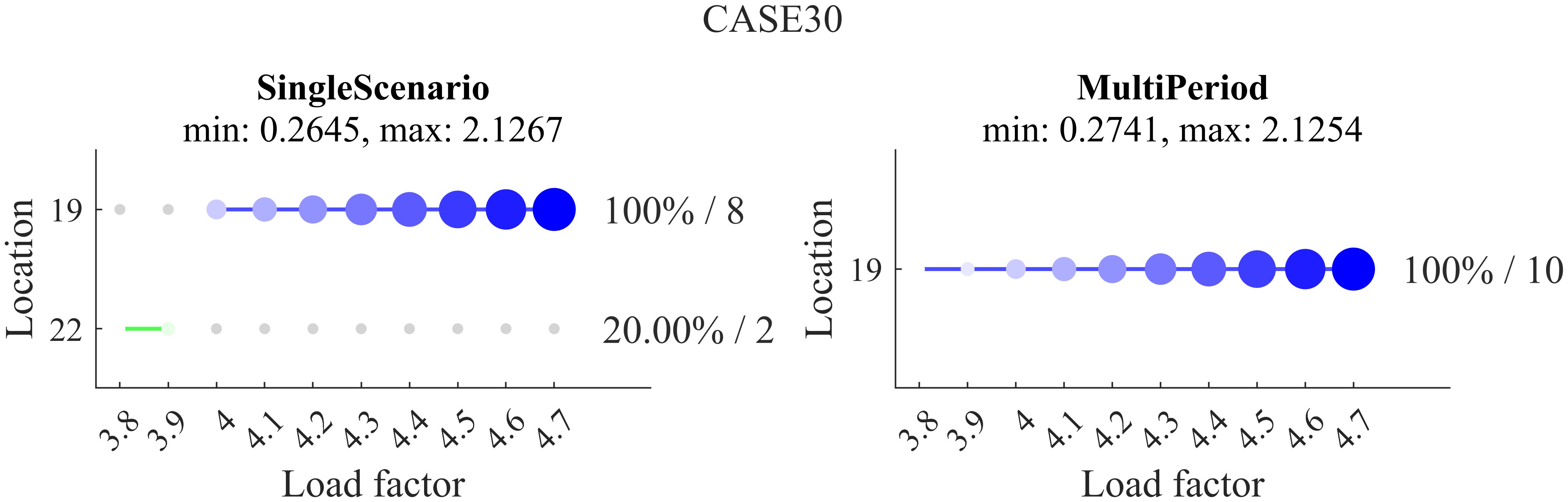}
         \caption{}
         \label{fig: persistency case30 a}
    \end{subfigure}
    \hfill
    \begin{subfigure}[b]{\linewidth}
         \centering       
         \includegraphics[width=\linewidth]{fig4-case30_setpersistency.jpg}
         \caption{}
         \label{fig: persistency case30 c}
     \end{subfigure}
    \caption[]{\textsc{Case30} Persistency: single-scenario (left) VS proposed multi-period (right) method. Subfigure (a) plots the key vulnerable locations at each scenario and plots the magnitude of key vulnerability sources $|n_i^{(t)}|$ by the size of the colored circle markers. Our proposed method (right) identifies bus \#19 as the only vulnerable location starting from 1st and persists until the last scenario; the location is a 100\% persistent failure location, i.e., the location persistency is 100\%, and it lasts for 10 steps in total (as notified on the right with ``100\% / 10''). Whereas the baseline method (left) identifies bus \#22 as the vulnerability source in the first two scenarios, it later switches to bus \#19, resulting in lower location persistency. The bottom plots (b) illustrate overall persistency by $\mathrm{SetPersistency}$ metric for baseline (left) and proposed (right) method. Bar plots show the number of locations $\vert\mathcal{S}^{(t)}\vert$ and the cumulative number of locations $\vert\cup_{j=1}^t \mathcal{S}^{(j)}\vert$ for $\forall\,t$; and red/blue line plot shows the ratio between the two, which is $\mathrm{SetPersistency(t)} = \vert\mathcal{S}^{(t)}\vert/\vert\cup_{j=1}^t \mathcal{S}^{(j)}\vert$ for $\forall\,t$.}
    \label{fig: persistency case30}
\end{figure}

\begin{figure}
\centering
\begin{subfigure}[b]{\linewidth}
     \centering         
     \includegraphics[width=\linewidth]{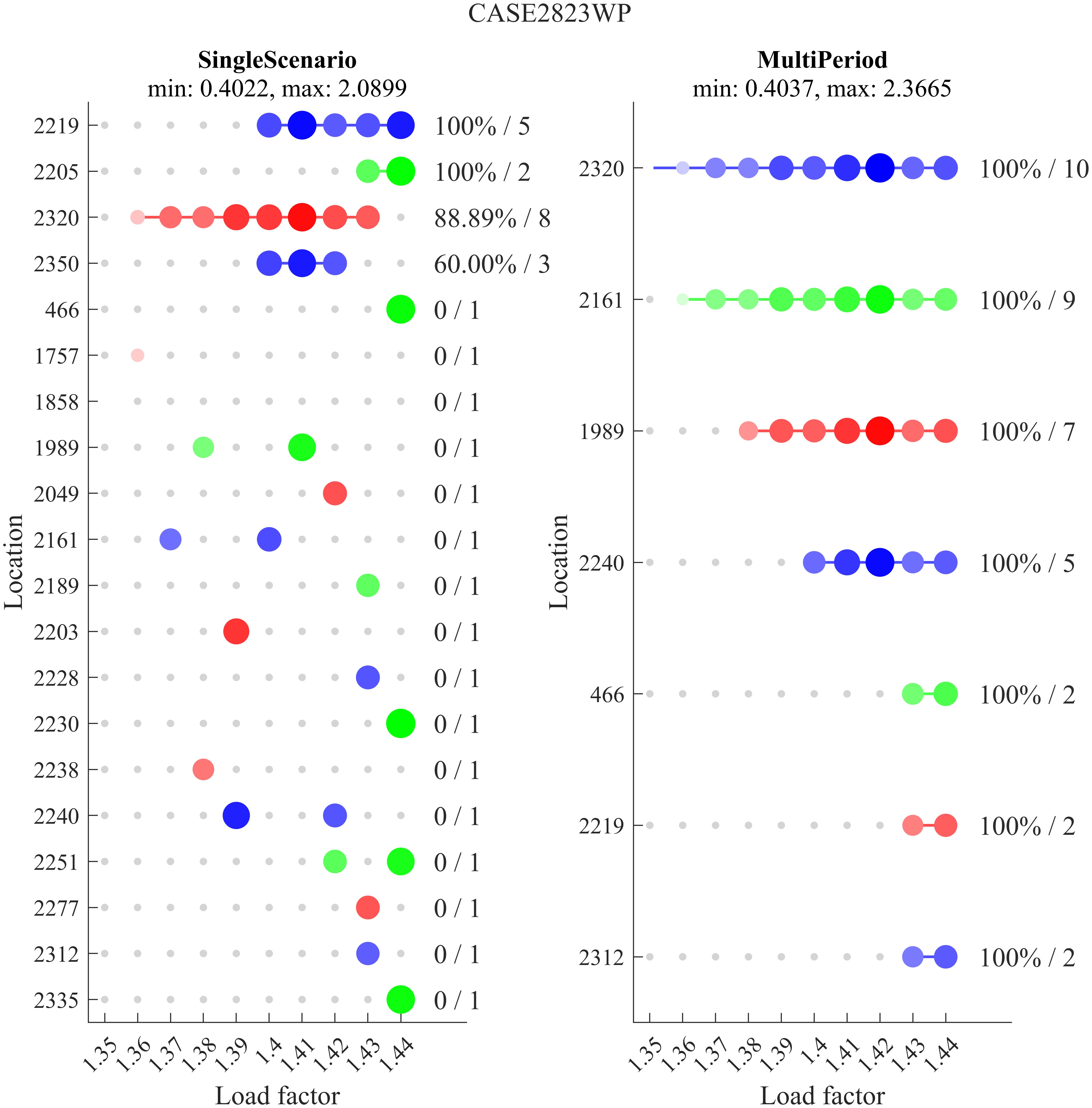}
     \caption{}
     \label{fig: persistency case30 a}
\end{subfigure}
\hfill
\begin{subfigure}[b]{\linewidth}
     \centering         
     \includegraphics[width=\linewidth]{fig5-case2383wp_setpersistency.jpg}
     \caption{}
     \label{fig: persistency case2383w c}
\end{subfigure}
\caption{\textsc{Case2383wp} Persistency: single-scenario (left) VS proposed multi-period (right) method. (a): Using single-scenario baseline, only 2 ideally persistent vulnerability locations (\#2219 and \#2205) are identified, all other vulnerable locations do not persist well, whereas using the proposed method, all identified vulnerability locations are persistent. (b): Set persistency decreases as more scenarios are processed using baseline method, whereas the proposed method maintains high set persistency.}
\label{fig: persistency case2383wp}
\end{figure}

Figure \ref{fig: persistency case30} and Figure \ref{fig: persistency case2383wp} show the persistency analysis on \textsc{case30} and \textsc{case2383wp} respectively. Results show that the proposed method achieves higher location persistency and set persistency than the baseline.

\subsection{Sparsity and total compensation} \label{sec: sparsity and total compensation}
The key goal of sparse diagnosis is to give targeted and focused insights of key vulnerable locations. In general, higher sparsity is more desirable.  So we evaluate whether the proposed method leads to less sparse solutions (i.e., more vulnerabilities) than the baseline, while encouraging persistency.

Meanwhile, the total amount of vulnerability sources represents intuitively the overall severity of the collapse and the total necessary compensation resources needed to restore system feasibility. Therefore, we evaluate whether the proposed method leads to a larger amount of total compensation resources, which could increase the risk of exaggerating severity or ``wasting'' resources. 

\begin{figure}
    \centering
    \begin{subfigure}[b]{0.49\linewidth}
        \centering
        \includegraphics[width=\linewidth]{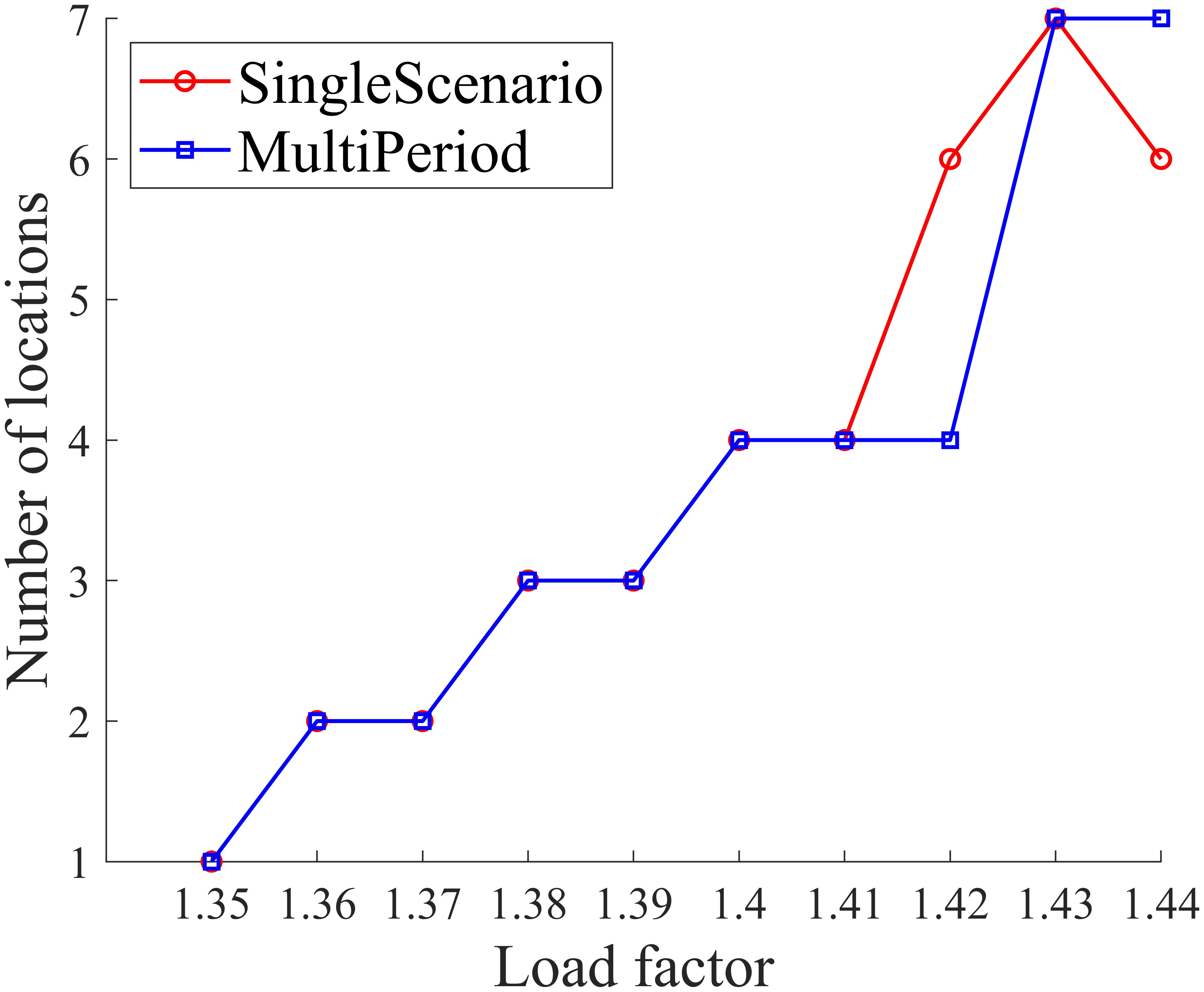}
        \caption{}
        \label{fig: total feasibility a}
    \end{subfigure} \hfill
    \begin{subfigure}[b]{0.49\linewidth}
        \centering
        \includegraphics[width=\linewidth]{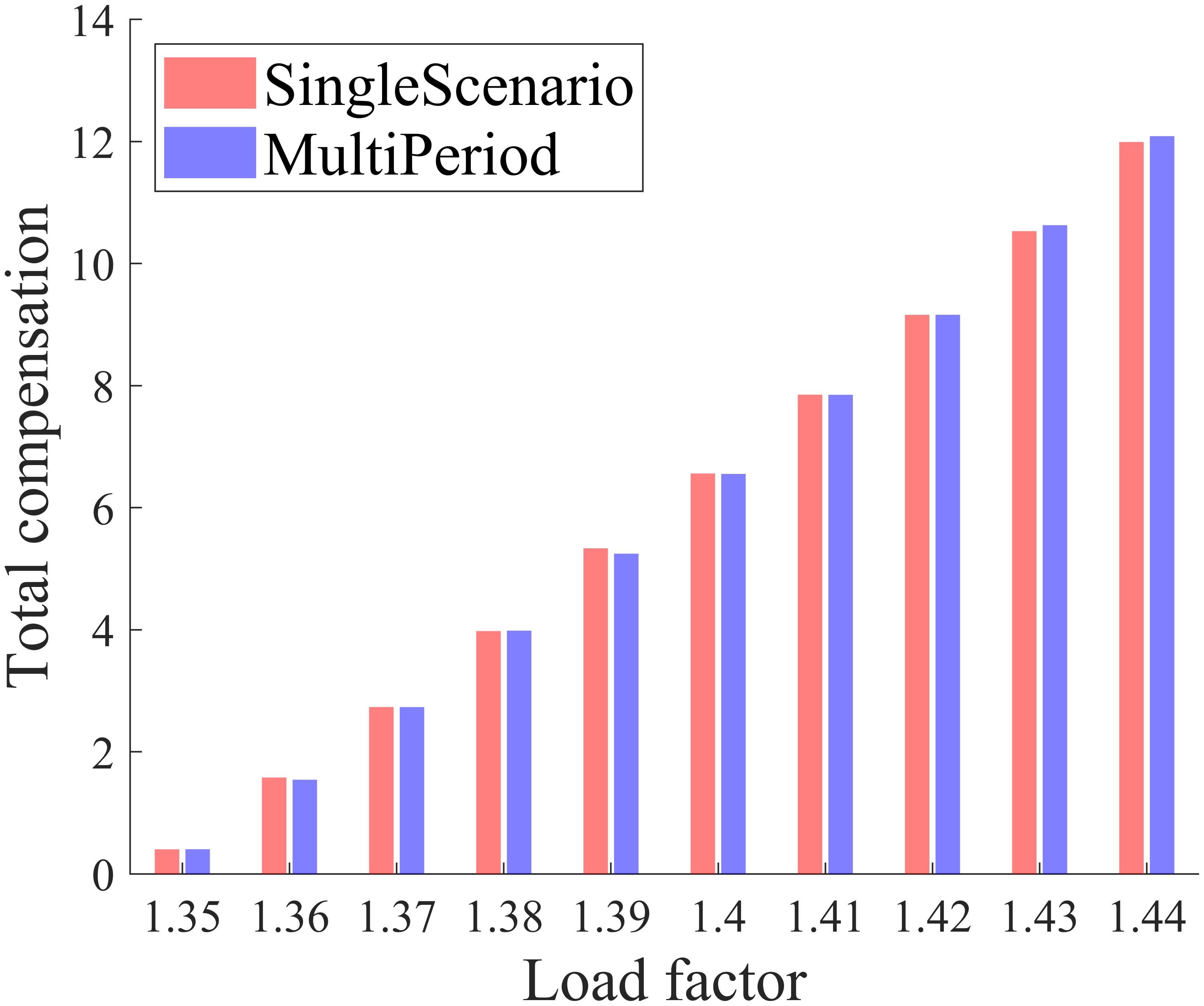}
        \caption{}
        \label{fig: total feasibility b}
    \end{subfigure}
    \caption{\textsc{Case2383wp}: Proposed multi-period method and the single-scenario baseline achieve comparable level of sparsity (a) and total amount of compensation sources (b) to restore feasibility, i.e., our proposed work promotes persistency on multiple scenario analysis without significantly sacrificing sparsity and compensation resources.}
    \label{fig: total feasibility}
\end{figure}

Figure \ref{fig: total feasibility} shows the results on \textsc{case2383wp}. Figure \ref{fig: total feasibility a} compares the sparsity, i.e. $|\mathcal{S}^{(t)}|$, of two methods, and shows that two methods are comparable, meaning that our proposed method does not significantly sacrifice sparsity when seeking to achieve the expected persistency. Figure \ref{fig: total feasibility b} also shows that the total compensation is not sacrificed.

Furthermore, from a practical perspective, a sparser solution together with a smaller total compensation generally means less additional power planning. A major advantage brought by persistency is to minimize intervention  not just for a single scenario, but for a time window of multiple scenarios. For example, when load demand increases at the next stage, it can be handled by slightly modifying the plan decision of the previous stage (e.g., making an already planned power plant bigger), instead of switching to a completely different plan.

\subsection{Scalability} \label{sec: scalability result}
To verify the scalability of the proposed method, we evaluate its run time on systems of different sizes. With similar experiment settings as above, we tested $10$ collapsed scenarios for each system by incrementally increasing their load factors within a range: \textsc{case30}\cite{matpower-case30-118-300-2383wp} (load factor 3.8$\sim$4.8), \textsc{case118}\cite{matpower-case30-118-300-2383wp} (3.5$\sim$4.5), \textsc{case300}\cite{matpower-case30-118-300-2383wp} (1.05$\sim$1.1), \textsc{case1354pegase}\cite{case1354pegase-1}\cite{case1354pegase-2} (1.32$\sim$1.35), \textsc{case2383wp}\cite{matpower-case30-118-300-2383wp} (1.35$\sim$1.44), and \textsc{case3375wp} (1.2$\sim$1.3).

\begin{figure}[htbp]
    \centering
    \includegraphics[width=0.5\textwidth]{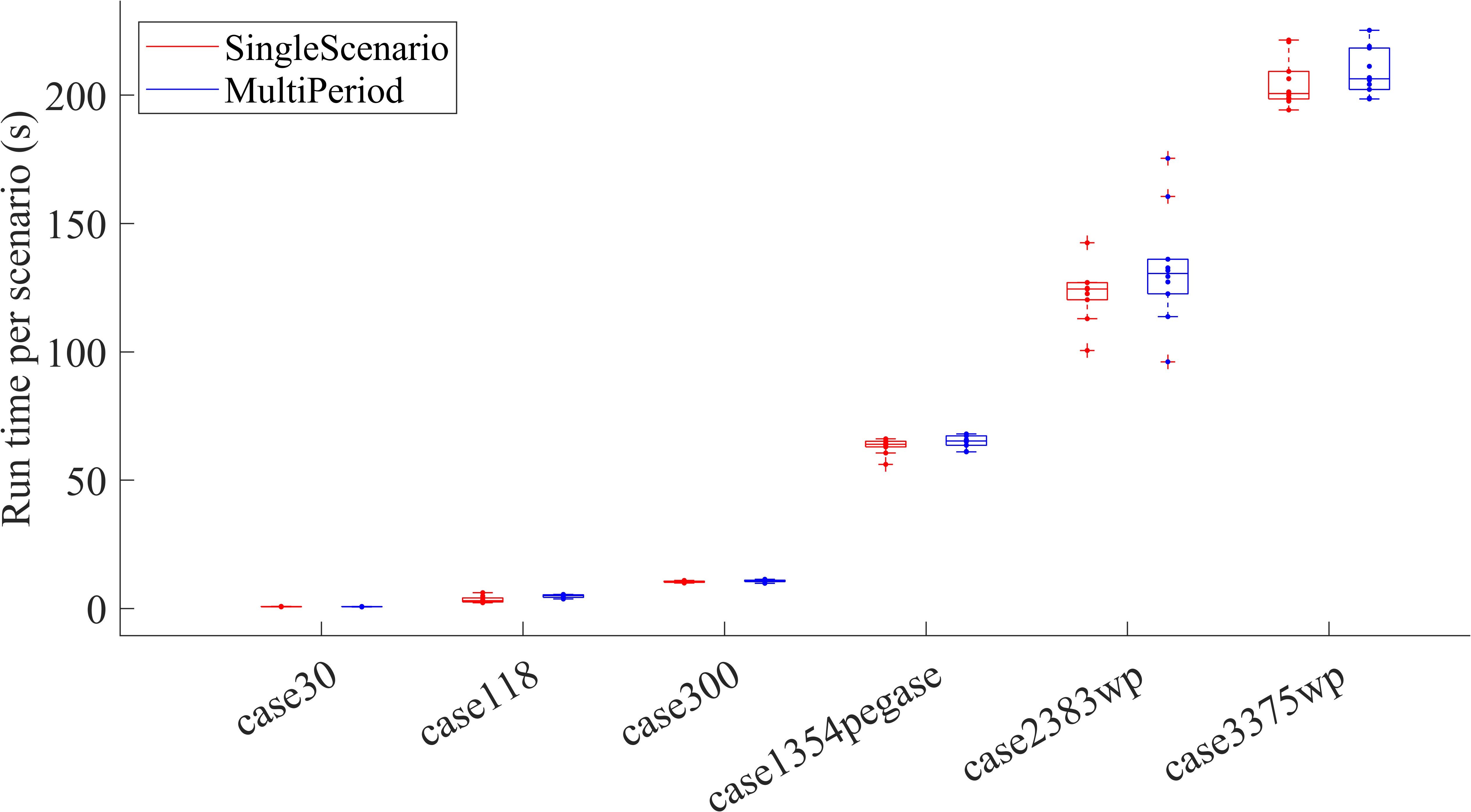}
    \caption{Comparison of per-scenario run time. Each scatter point represents the run time of an individual scenario in the test case, while the corresponding  box plot summarizes the overall distribution, including the median, 25th and 75th percentiles, and the minimum and maximum values.
    \label{fig: scalability}}
\end{figure}

Figure \ref{fig: scalability} shows that our proposed method scales well on large-scale systems, and the run time is comparable with the single-scenario baseline. This verifies that our multi-period approach encourages persistency in an efficient way without introducing significant computation burden. Moreover, it is worth noting that analyzing power flows under various load growth scenarios (the use case being considered here) is primarily a system planning application. In this context, the capability to solve for systems with thousands of buses in just around 3-4 minutes is highly promising and shows the computational robustness of the proposed methodology.

\subsection{Extended experiments on uneven load-growth cases} \label{sec: extended experiments} 
In this subsection, we extend the experimentation on uneven load-growth, specifically considering two heterogeneous patterns:
\subsubsection{Location-heterogeneous load growth} Each individual load location experiences a different random stress level. At each time step $t$, a random load factor $\text{LF}_i(t)$ is assigned to each load bus $i$ by sampling from a uniform distribution $[70\%, 130\%]\cdot\overline{\text{LF}}(t)$, with the base (average) load factor $\overline{\text{LF}}(t)$ increasing over time. 

\subsubsection{Area-heterogeneous load growth} Loads are grouped into subsets corresponding to different zones/areas, and each zone/area experiences a different stress level. At each time step, we assign a random load factor $\text{LF}_\text{area $j$}(t)$ to loads with the same area $j$, while enforcing the total load increase to be at a base (average) level $\overline{\text{LF}}(t)$. To this end, we adopt a weighted random sampling, by first sampling area-specific values from the normal distribution $\xi_j\sim\mathcal{N}(0,1)$ and then assign:
\begin{equation}
    \text{LF}_\text{area $j$}(t) = \dfrac{\mathrm{e}^{\sigma\xi_j}}{\sum_{k=1}^m w_k\cdot\mathrm{e}^{\sigma\xi_k}}\cdot\overline{\text{LF}}(t)
\end{equation}
where the area weight is defined as:
\begin{equation}
    w_j = \dfrac{P_\text{area $j$}}{\sum_{k=1}^m P_\text{area $k$}}.
\end{equation}
$P_\text{area $j$}$ denotes the total real power in area $j$, and the hyperparameter $\sigma$ controls heterogeneity intensity (we set $\sigma = 0.05$). Larger $\sigma$ creates bigger differences in $\text{LF}_\text{area j}(t)$ across areas. 

\begin{figure}
    \centering
    \begin{subfigure}[b]{\linewidth}
        \centering
        \includegraphics[width=\linewidth]{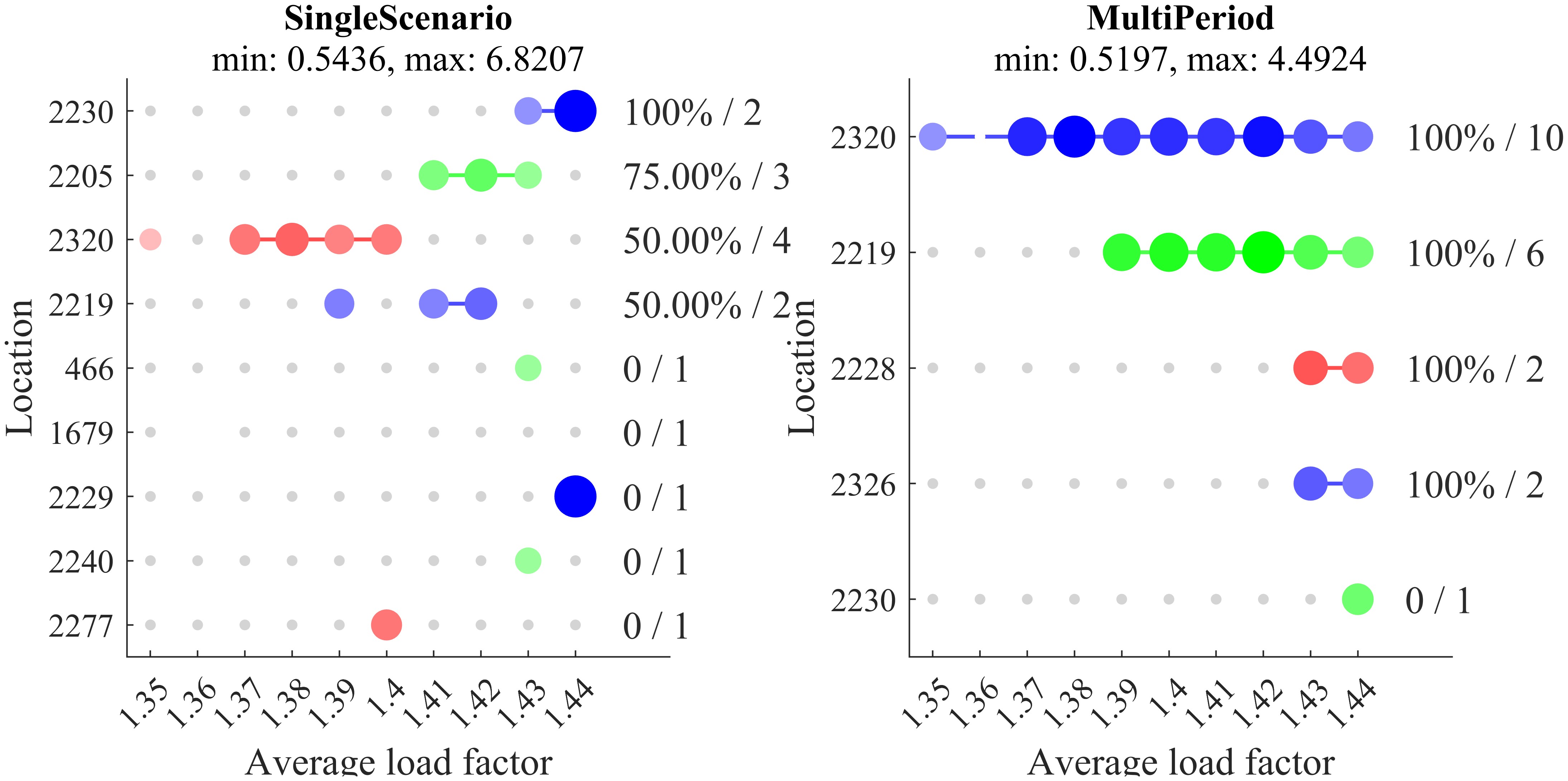}
        \caption{}
    \end{subfigure} \\
    \begin{subfigure}[b]{\linewidth}
        \centering
        \includegraphics[width=\linewidth]{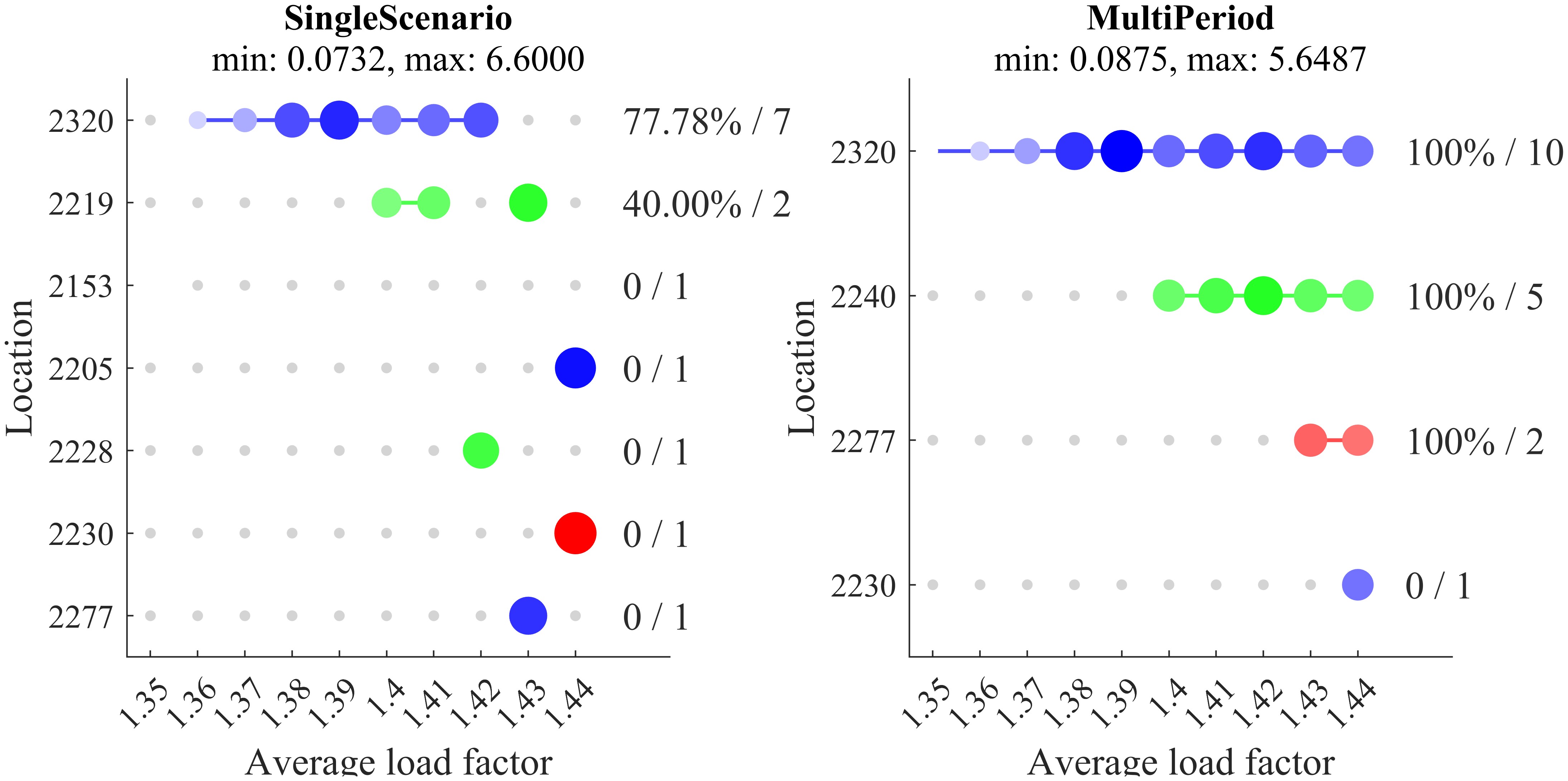}
        \caption{}
    \end{subfigure}
    \caption{\textsc{case2383wp} Persistency under uneven load growth:  single-scenario (left) VS multi-period (right) method. In the  experiments: (a) location-heterogeneous and (b) area-heterogeneous load growth are evaluated.}
    \label{fig: extended experiments}
\end{figure}

On \textsc{case2383wp}, Figure \ref{fig: extended experiments} compares the persistency performance between the single-scenario baseline and the proposed multi-period method. The results indicate that the multi-period method effectively identifies persistent vulnerability locations despite spatially heterogeneous load growth.

\section{Discussion and future work}

A key demonstrated application of this paper is to long-term planning studies in which the sequence of $T$ scenarios is strongly and temporally correlated by increasing stress driven by increasing load demand (both evenly and unevenly). The idea extends, without loss of generality, to cascading failure analysis where the scenarios are driven by the growing stress of worsening contingencies, or deteriorating instability, which we would like to reserve for our future work. 

Specifically, when evaluating cascading failures, advanced methods of cascading outage simulations \cite{DC-cascading-transient, bienstock2010optimal} or transient analysis \cite{wu2023transient, transient-analysis} can reveal progression toward collapse by starting from the system before disturbance and solving a series of feasible power flow problems to determine the worsening operating states sequentially before the system collapses. However, these methods are in general not able to solve the final collapsed case and thus give limited insights into dominant failure sources for the eventual collapsed state. In this context, our future work aims to advance the existing studies, by applying slack compensation sources to capture the evolving and deteriorating vulnerabilities behind the spread of failures. In practical decision-making, a trajectory of sparse compensations potentially indicates the targeted corrective actions to restore the stable normal operating point at different stages of the cascading failure propagation.

Furthermore, beyond the strongly correlated scenarios analyzed in temporal planning and cascading failures, planners and operators in practice also routinely study many scenarios that are not temporally correlated, such as a set of N-1 line outages at different locations \cite{carrion2021novel,glanzmann2006incorporation,chakrabarti2014security}. Existing contingency-analysis workflows typically evaluate post-contingency states scenario by scenario, and cannot converge when any contingency leads to blackout. For such applications, we aim to advance contingency analysis by co-simulating multiple scenarios to identify the shared vulnerabilities behind them, which, in practice, potentially indicate placing of active and/or reactive power resources (e.g. FACTS devices) to make the system N-1 secure.

\section{Conclusion}
\label{sec:Conclusion}
This paper proposes a novel multi-period sparse optimization to identify the vulnerable locations responsible for multiple power system blackouts induced by extremely high demand conditions. Compared with analyzing each scenario separately, our multi-period method sorts all scenarios in the order of growing stress to form a sequence of multi-period practical load-growth scenarios, and then encourages persistency in vulnerability identification problem while solving for these cases. We defined the ideal persistency of vulnerability sources, and two evaluation metrics: locational persistency and set persistency. Experiments on benchmark systems show that the method 1) reliably tracks {persistent failure locations} under increasing load stress, pinpointing 7 persistent vulnerability locations on \textsc{Case2383wp} with the range of 35\%$\sim$44\% load increase; 2) solves with scalability to large systems (taking $<4$ min on 3000+ bus systems). The proposed work itself is agnostic to demand patterns and has demonstrated effectiveness under both uniformly growing load factors and uneven load growth, without loss of generality. Moreover, by capturing inherent vulnerabilities across a continuous stress spectrum, this work enables projections of intermediate load-growth scenarios that are not explicitly solved. In planning studies, this capability reduces the need to evaluate every critical case individually, thereby significantly improving efficiency.

\bibliographystyle{IEEEtran}
\bibliography{pscc2025}

@article{2021TexasPowerCrisis,
  title={Cascading risks: Understanding the 2021 winter blackout in Texas},
  author={Busby, Joshua W and Baker, Kyri and Bazilian, Morgan D and Gilbert, Alex Q and Grubert, Emily and Rai, Varun and Rhodes, Joshua D and Shidore, Sarang and Smith, Caitlin A and Webber, Michael E},
  journal={Energy Research \& Social Science},
  volume={77},
  pages={102106},
  year={2021},
  publisher={Elsevier}
}

@article{ukrain2015,
  title={Analysis of the cyber attack on the Ukrainian power grid},
  author={Case, Defense Use},
  journal={Electricity Information Sharing and Analysis Center (E-ISAC)},
  volume={388},
  year={2016}
}

@article{lee2017crashoverride,
  title={CRASHOVERRIDE: Analysis of the threat to electric grid operations},
  author={Lee, Robert M and Assante, MJ and Conway, T},
  journal={Dragos Inc., March},
  year={2017}
}

@inproceedings{madiot,
  title={$\{$BlackIoT$\}$:$\{$IoT$\}$ botnet of high wattage devices can disrupt the power grid},
  author={Soltan, Saleh and Mittal, Prateek and Poor, H Vincent},
  booktitle={27th USENIX security symposium (USENIX security 18)},
  pages={15--32},
  year={2018}
}

@inproceedings{SparseFeas,
  title={A lasso-inspired approach for localizing power system infeasibility},
  author={Li, Shimiao and Pandey, Amritanshu and Agarwal, Aayushya and Jereminov, Marko and Pileggi, Larry},
  booktitle={2020 IEEE Power \& Energy Society General Meeting (PESGM)},
  pages={1--5},
  year={2020},
  organization={IEEE}
}

@article{powerblackout,
  title={The anatomy of a power grid blackout-root causes and dynamics of recent major blackouts},
  author={Pourbeik, Pouyan and Kundur, Prabha S and Taylor, Carson W},
  journal={IEEE Power and Energy Magazine},
  volume={4},
  number={5},
  pages={22--29},
  year={2006},
  publisher={IEEE}
}

@article{powerflow-diverge,
  title={The continuation power flow: a tool for steady state voltage stability analysis},
  author={Ajjarapu, Venkataramana and Christy, Colin},
  journal={IEEE transactions on Power Systems},
  volume={7},
  number={1},
  pages={416--423},
  year={1992},
  publisher={Ieee}
}

@article{sugar-pf,
  title={Evaluating feasibility within power flow},
  author={Jereminov, Marko and Bromberg, David M and Pandey, Amritanshu and Wagner, Martin R and Pileggi, Larry},
  journal={IEEE Transactions on Smart Grid},
  volume={11},
  number={4},
  pages={3522--3534},
  year={2020},
  publisher={IEEE}
}

@article{infeasibility-quantified-pf-trad,
  title={A power flow measure for unsolvable cases},
  author={Overbye, Thomas J},
  journal={IEEE Transactions on Power Systems},
  volume={9},
  number={3},
  pages={1359--1365},
  year={1994},
  publisher={IEEE}
}

@inproceedings{sparsePFref5-planningFACTS4congestion,
  author = {Manasarani Mandala and C. P. Gupta},
  title = {Congestion management by optimal placement of FACTS device},
  booktitle = {2010 Joint International Conference on Power Electronics, Drives and Energy Systems \& 2010 Power India},
  year = {2010},
  publisher = {IEEE}
}

@article{sparsePFref6-planningSVCnTCSC,
  author = {R. Benabid and M. Boudour and M. A. Abido},
  title = {Optimal location and setting of SVC and TCSC devices using non-dominated sorting particle swarm optimization},
  journal = {Electric Power Systems Research},
  volume = {79},
  number = {12},
  pages = {1668--1677},
  year = {2009}
}

@article{sparsePFref7-planningSVC,
  author = {Malihe M. Farsangi and H. Nezamabadi-pour and Y. H. Song and K. Y. Lee},
  title = {Placement of SVCs and selection of stabilizing signals in power systems},
  journal = {IEEE Transactions on Power Systems},
  volume = {22},
  number = {3},
  pages = {1061--1071},
  year = {2007}
}

@article{sparsePFref8-planningFACTS,
  author = {Nikhlesh Kumar Sharma and Arindam Ghosh and Rajiv Kumar Varma},
  title = {A novel placement strategy for FACTS controllers},
  journal = {IEEE Transactions on Power Delivery},
  volume = {18},
  number = {3},
  pages = {982--987},
  year = {2003}
}

@article{sparsePFref12-SUGAR-opf,
  author = {Marko Jereminov and Amritanshu Pandey and Larry Pileggi},
  title = {Equivalent Circuit Formulation for Solving AC Optimal Power Flow},
  journal = {IEEE Transactions on Power Systems},
  volume = {34},
  number = {3},
  pages = {2354--2365},
  year = {2018}
}

@book{sparsePFref15-spice,
  author = {Paul W. Tuinenga},
  title = {SPICE: a guide to circuit simulation and analysis using PSpice},
  volume = {2},
  publisher = {Prentice Hall},
  year = {1995}
}

@article{sugar-pf-Amrit,
  title={Robust power flow and three-phase power flow analyses},
  author={Pandey, Amritanshu and Jereminov, Marko and Wagner, Martin R and Bromberg, David M and Hug, Gabriela and Pileggi, Larry},
  journal={IEEE Transactions on Power Systems},
  volume={34},
  number={1},
  pages={616--626},
  year={2018},
  publisher={IEEE}
}

@inproceedings{grid-security-Vyas,
  title={Shedding Light on Inconsistencies in Grid Cybersecurity: Disconnects and Recommendations},
  author={Singer, Brian and Pandey, Amritanshu and Li, Shimiao and Bauer, Lujo and Miller, Craig and Pileggi, Lawrence and Sekar, Vyas},
  booktitle={2023 IEEE Symposium on Security and Privacy (SP)},
  pages={554--571},
  year={2022},
  organization={IEEE Computer Society}
}

@article{SparseFeas4DNet,
  title={Three-phase infeasibility analysis for distribution grid studies},
  author={Foster, Elizabeth and Pandey, Amritanshu and Pileggi, Larry},
  journal={Electric Power Systems Research},
  volume={212},
  pages={108486},
  year={2022},
  publisher={Elsevier}
}

@article{SparseFeas4TnDNet,
  title={Distributed Primal-Dual Interior Point Framework for Analyzing Infeasible Combined Transmission and Distribution Grid Networks},
  author={Ali, Muhammad Hamza and Pandey, Amritanshu},
  journal={arXiv preprint arXiv:2409.14532},
  year={2024}
}

@inproceedings{SUGAR-SE-Li,
  author={Li, Shimiao and Pandey, Amritanshu and Kar, Soummya and Pileggi, Larry},
  booktitle={2020 IEEE PES Innovative Smart Grid Technologies Europe (ISGT-Europe)}, 
  title={A Circuit-Theoretic Approach to State Estimation}, 
  year={2020},
  volume={},
  number={},
  pages={1126-1130},
  doi={10.1109/ISGT-Europe47291.2020.9248872}
}

@article{convexSE-LAV-Li,
  title={A WLAV-based Robust Hybrid State Estimation using Circuit-theoretic Approach},
  author={Li, Shimiao and Pandey, Amritanshu and Pileggi, Larry},
  journal={arXiv preprint arXiv:2011.06021},
  year={2020}
}

@article{ckt-GSE,
  title={A convex method of generalized state estimation using circuit-theoretic node-breaker model},
  author={Li, Shimiao and Pandey, Amritanshu and Pileggi, Larry},
  journal={arXiv preprint arXiv:2109.14742},
  year={2021}
}

@incollection{ckt-SE-TnD,
  title={Combined transmission and distribution state-estimation for future electric grids},
  author={Pandey, Amritanshu and Li, Shimiao and Pileggi, Larry},
  booktitle={Power Systems Operation with 100\% Renewable Energy Sources},
  pages={299--315},
  year={2024},
  publisher={Elsevier}
}

@article{gridwarm,
  title={Contingency analysis with warm starter using probabilistic graphical model},
  author={Li, Shimiao and Pandey, Amritanshu and Pileggi, Larry},
  journal={Electric Power Systems Research},
  volume={234},
  pages={110737},
  year={2024},
  publisher={Elsevier}
}

@article{matpower-case30-118-300-2383wp,
  title={MATPOWER: Steady-state operations, planning, and analysis tools for power systems research and education},
  author={Zimmerman, Ray Daniel and Murillo-S{\'a}nchez, Carlos Edmundo and Thomas, Robert John},
  journal={IEEE Transactions on power systems},
  volume={26},
  number={1},
  pages={12--19},
  year={2010},
  publisher={IEEE}
}

@article{case1354pegase-1,
  title={AC power flow data in MATPOWER and QCQP format: iTesla, RTE snapshots, and PEGASE},
  author={Josz, C{\'e}dric and Fliscounakis, St{\'e}phane and Maeght, Jean and Panciatici, Patrick},
  journal={arXiv preprint arXiv:1603.01533},
  year={2016}
}

@article{case1354pegase-2,
  title={Contingency ranking with respect to overloads in very large power systems taking into account uncertainty, preventive, and corrective actions},
  author={Fliscounakis, St{\'e}phane and Panciatici, Patrick and Capitanescu, Florin and Wehenkel, Louis},
  journal={IEEE Transactions on Power Systems},
  volume={28},
  number={4},
  pages={4909--4917},
  year={2013},
  publisher={IEEE}
}

@article{bienstock2010optimal,
  title={Optimal adaptive control of cascading power grid failures},
  author={Bienstock, Daniel},
  journal={arXiv preprint arXiv:1012.4025},
  year={2010}
}

@article{wu2023transient,
  title={Transient stability analysis of large-scale power systems: A survey},
  author={Wu, Qing-Hua and Lin, Yuqing and Hong, Chao and Su, Yinsheng and Wen, Tianhao and Liu, Yang},
  journal={CSEE Journal of Power and Energy Systems},
  volume={9},
  number={4},
  pages={1284--1300},
  year={2023},
  publisher={CSEE}
}

@article{DC-cascading-transient,
  title={Cascading failure analysis with DC power flow model and transient stability analysis},
  author={Yan, Jun and Tang, Yufei and He, Haibo and Sun, Yan},
  journal={IEEE Transactions on Power Systems},
  volume={30},
  number={1},
  pages={285--297},
  year={2014},
  publisher={IEEE}
}

@book{transient-analysis,
  title={Power system transient stability analysis using the transient energy function method},
  author={Fouad, Abdel-Azia and Vittal, Vijay},
  year={1991},
  publisher={Pearson Education}
}

@article{carrion2021novel,
  title={A novel methodology for optimal svc location considering n-1 contingencies and reactive power flows reconfiguration},
  author={Carri{\'o}n, Diego and Garc{\'\i}a, Edwin and Jaramillo, Manuel and Gonz{\'a}lez, Jorge W},
  journal={Energies},
  volume={14},
  number={20},
  pages={6652},
  year={2021},
  publisher={MDPI}
}

@inproceedings{glanzmann2006incorporation,
  title={Incorporation of n-1 security into optimal power flow for facts control},
  author={Glanzmann, Gabriela and Andersson, Goran},
  booktitle={2006 IEEE PES power systems conference and exposition},
  pages={683--688},
  year={2006},
  organization={IEEE}
}

@inproceedings{chakrabarti2014security,
  title={Security constrained optimal power flow via proximal message passing},
  author={Chakrabarti, Sambuddha and Kraning, Matt and Chu, Eric and Baldick, Ross and Boyd, Stephen},
  booktitle={2014 Clemson university power systems conference},
  pages={1--8},
  year={2014},
  organization={IEEE}
}

\balance

\end{document}